\documentclass{article}
\usepackage[dvips]{graphicx}
\usepackage{color}

\begin{document}

% waiting line primitive
\newsavebox{\waitline}
\savebox{\waitline}(0,0){
	\put( 5,20){\vector(1,0){16}}
	\put(25,10){\framebox(10,20)}
	\put(37,10){\framebox(10,20)}
	\put(49,10){\framebox(10,20)}
	\put(59,20){\line(1,0){6}}
}

% single server primitive
\newsavebox{\server}
\savebox{\server}(0,0){
	\put(59,18){\line(1,0){6}}
	\put(75,18){\circle{20}}
	\put(72,15){D}
	\put(90,18){\vector(1,0){16}}
}

% Coxian server primitive
\newsavebox{\coxian}
\savebox{\coxian}(0,0){

\put(0,-2){\dashbox(150,40)[l]{

	\put(-10,18){\vector(1,0){15}}
	\put(-10,25){$a_{1}$}
	\put(15,18){\circle{20}}
	\put(10,15){$\mu_{1}$}	
	\put(25,18){\vector(1,0){15}}
	\put(27,25){$a_{2}$}
	\put(25,18){\vector(0,-1){20}}
	
	\put(50,18){\circle{20}}
	\put(45,15){$\mu_{2}$}	
	\put(60,18){\vector(1,0){15}}
	\put(60,18){\vector(0,-1){20}}

	\put(80,18){$\ldots$}

	\put(100,18){\vector(1,0){15}}
	\put(100,25){$a_{p}$}
	\put(125,18){\circle{20}}
	\put(120,15){$\mu_{p}$}	
	\put(135,18){\vector(0,-1){20}}
	
	\put(25,-2){\vector(1,0){50}}
	\put(80,-2){$\ldots$}
	\put(100,-2){\vector(1,0){60}}
}}

}

% think server primitive
\newsavebox{\think}
\savebox{\think}(0,0){
	\put(59,30){\vector(1,1){15}}
	\put(59,30){\vector(1,0){12}}
	\put(59,30){\vector(1,-1){15}}
	
	\put(85,57){\circle{15}}
	\put(83,54){Z}
	\put(85,37){\circle{15}}
	\put(83,34){Z}
	\put(83,15){\vdots}
	\put(85,3){\circle{15}}
	\put(83,0){Z}
	\put(100,30){\vector(1,0){30}}
	
	\put(95,45){\line(1,-1){15}}
	\put(95,30){\line(1,0){12}}
	\put(95,15){\line(1,1){15}}
}

% M/M/1 primitive
\newsavebox{\MMuno}
\savebox{\MMuno}(0,0){
	\put(0,0){\usebox{\waitline}}
	\put(5,1){\usebox{\server}}
}

\title{A New Interpretation of Amdahl's Law and Geometric Scaling 
\thanks{Copyright ~\copyright ~ 2002 Performance Dynamics Company.
All Rights Reserved.}}
\author{Neil J. Gunther 
\thanks{Email correspondence: \tt njgunther@perfdynamics.com} \\ \\
\small \it Performance Dynamics Company, Castro Valley, CA 94552, U.S.A.\rm
}
\date{\small Draft of April 19, 2002}

\maketitle

\begin{abstract}
The multiprocessor effect refers to the loss of computing cycles
due to processing overhead. Amdahl's law and the Multiprocessing
Factor (MPF) are two scaling models used in industry and academia
for estimating multiprocessor capacity in the presence of this
multiprocessor effect. Both models express different laws of
diminishing returns. Amdahl's law identifies diminishing
processor capacity with a fixed degree of serialization in the
workload, while the MPF model treats it as a constant geometric
ratio. The utility of both models for performance evaluation
stems from the presence of a single parameter that can be
determined easily from a small set of benchmark measurements.
This utility, however, is marred by a dilemma. The two models
produce different results, especially for large processor
configurations that are so important for today's applications.
The question naturally arises: Which of these two models is the
correct one to use?  Ignoring this question merely reduces
capacity prediction to arbitrary curve-fitting. Removing the
dilemma requires a dynamical interpretation of these scaling
models. We present a physical interpretation based on queueing
theory and show that Amdahl's law corresponds to synchronous
queueing in a bus model while the MPF model belongs to a Coxian
server model. The latter exhibits unphysical effects such as
sublinear response times hence, we caution against its use for
large multiprocessor configurations. 
\end{abstract}

\section{Introduction}
The multiprocessor effect is a generic term for the fraction of processing
cycles usurped by the system (both software and hardware) in order to execute a given workload.  
Typical sources of multiprocessor overhead include:
\begin{enumerate}
\item	Operating system code paths (system calls in Unix; supervisor calls in MVS)
\item	Exchange of shared writable data between processor caches across the system bus
\item	Data exchange between processors across the system bus to main memory
\item	Lock synchronization of accesses to shared writable data 
\item	Waiting for a I/O to complete
\end{enumerate}
In the absence of such overhead, the aggregate processor capacity would scale linearly. This
could occur if there were single-threaded applications running on each processor.
More commonly, however, diminishing processing capacity reduces the potential economies of 
scale offered by symmetric multiprocessors; a point first observed by Gene Amdahl~\cite{amdahl}. 

The multiprocessor effect can be viewed as a type of interaction between processors as they contend 
for shared subsystem resources.  As more processors are added to the backplane (to process more 
work presumably) system overhead increases due to the increasing degree of processor interaction.  
This interaction exhibits itself as incremental capacity falling short of the linear ideal.
Therefore, any attempt to predict the multiprocessor effect requires a nonlinear function.

We examine a class of single parameter functions used for estimating multiprocessor capacity in the
presence of the multiprocessor effect.
For sizing p processors, the capacity functions C(p) must satisfy the following general criteria:
\begin{enumerate}
\item Concave function of p.
\item Monotonically increasing.
\item Vanishes at zero capacity: C(0) = 0.
\item Bounded above: $C(p) \rightarrow $ const. as $p \rightarrow \infty$.
\end{enumerate}

Two members of this class of capacity functions 
that have widespread application in industry and  
perennial discussion in the literature are: (i) Amdahl's law~\cite{amdahl}
\begin{equation}
C(\sigma, p) = \frac{p}{1 + \sigma ~(p - 1)}  \label{eqn:amdsize}
\end{equation}
commonly associated with parallel processors 
(\cite{patterson},~\cite{gelenbe},~\cite{ware}),
and (ii) the Multiprocessing Factor (MPF)
\begin{equation}
C(\phi, p) = \frac{~1 - \phi^{p}}{1 -\phi} \label{eqn:mpfsize}
\end{equation}
used for sizing multiprocessor platforms (\cite{artis},~\cite{cockwalk}, \cite{mcgall}, \cite{gsa});
particularly mainframe vendors (\cite{ibm}, ~\cite{hitachi}, ~\cite{unisys}).
In both (\ref{eqn:amdsize}) and (\ref{eqn:mpfsize}), the respective parameters $\sigma$ and $\phi$ 
are real-valued on the open interval $(0, 1)$.

\begin{figure}[!hbtp]
\begin{center} 
\includegraphics[bb = 0 0 288 177, scale = 0.75]{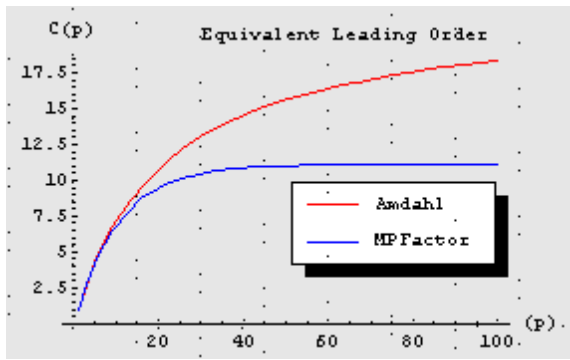} 
\caption{Common leading behaviour.} ~\label{fig:leading}
\end{center}
\end{figure}

Both equations express laws of diminishing returns (\ref{fig:leading}) but they
should not be regarded as laws in the sense of Little's law, however, because
they are not universal. Rather, they reflect a particular set of ad hoc
assumptions which we shall examine more closely in section \ref{adhoc}.

To set the perspective for what follows, we contrast the ad hoc application of 
(\ref{eqn:amdsize}) and (\ref{eqn:mpfsize}) to multiprocessor sizing with a more principled
methodology for sizing a memory or network buffer. 
Like $C(p)$, the buffer size,
$Q(\rho)$, belongs to a class of functions that must satisfy similar general criteria:
\begin{enumerate}
\item Be a convex function.
\item Be monotonically increasing on the interval $[0, 1]$.
\item Vanishes at zero load: $Q(0) = 0$.
\item Unbounded above: $Q(\rho) \rightarrow \infty $ as $\rho \rightarrow 1$.
\end{enumerate}
The queueing characteristics of different buffer models will have similar but
not identical curves (Fig. XXXXX).
In the process of characterizing the buffer size, one first selects a queueing
model (e.g., M/M/1 or M/G/1) based on an understanding of the buffer dynamics and
then validates the corresponding queue length formula $Q(\rho)$ against measurements. Even if this 
methodology is not strictly adhered to on every occasion, one has the option of doing it
this way.

Picking either of the processor sizing equations (\ref{eqn:amdsize}) and
(\ref{eqn:mpfsize}), on the other hand, is analogous to blindly choosing an ad hoc
queue length formula without any regard for the underlying queueing dynamics.
In this sense, it might be more accurate to refer to (\ref{eqn:amdsize}) and (\ref{eqn:mpfsize}) as
\emph{lores} for diminishing returns.

On the other hand, the usefulness of sizing equations like (\ref{eqn:amdsize})
and (\ref{eqn:mpfsize}) lies in the fact that there is only one parameter and it
can be determined easily by linear regression on just a few benchmark
measurements (\cite{lsprRB}, \cite{spec}, \cite{tpcc}, \cite{tpcw}). 
The question arises: Which is the best choice of parametric model
against which to fit the data?

If we assume the leading order characteristics are the same for small multiprocessor configurations,
Figure~\ref{fig:leading} shows that their respective asymptotes are very different and therefore
the parametric models predict very different large-scale configuration capacities.
Note that the MPF model saturates before the Amdahl model under these conditions and it therefore
predicts a smaller overall capacity than the Amdahl model.

\begin{figure}[!hbtp]
\begin{center} 
\includegraphics[bb = 0 0 288 177, scale = 0.75]{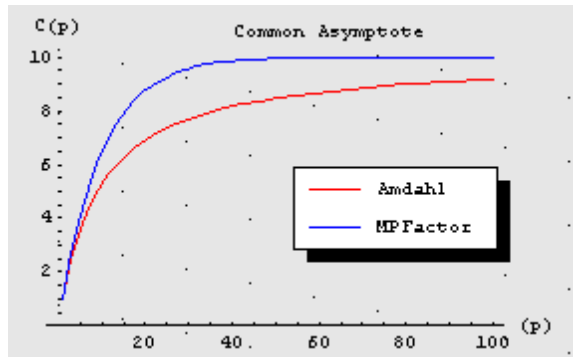} 
\caption{Common asymptote.} ~\label{fig:asymptotia}
\end{center}
\end{figure}

Alternatively, we could consider the other extreme shown in Figure~\ref{fig:asymptotia}
where both models approach the same asymptote at
for large multiprocessor configurations.
Now, the faster saturation of the MPF model means that maximal capacity is reached at
smaller configurations than predicted by the Amdahl model.

We take the position that such questions should be addressed on purely physical
grounds, otherwise, multiprocessor capacity predictions are reduced 
to an exercise in mere curve fitting.
The problem is that no consistent physical interpretation of these parametric
models exists.

Elsewhere~\cite{njgCMG96}, this author has shown how these parametric scaling models could be
expanded as a finite series in which each term has a distinct
pictorial representation. This led to the conclusion that
(\ref{eqn:amdsize}) can be regarded as representing a ``broadcast'' protocol while
(\ref{eqn:mpfsize}) can be regarded as representing a ``bucket brigade'' protocol~\cite{njgBOOK00}.
The latter is a less than satisfactory because it appears quite unphysical when
compared to the way actual multiprocessor systems operate.

In this paper, we present a more consistent interpretation based on queueing models. 
The usual difficulty with modeling multiprocessors as elementary queues (e.g., M/M/m) is that 
they do not account for ``interference'' effects between the processors
(the so-called \emph{multiprocessor effect}). 
We shall overcome this limitation in two distinct ways:
\begin{enumerate}
\item Multiprocessor Speedup will be identified with a bus-oriented M/M/1//p queueing
model where processor interference is represented as communication delays
across the bus.
\item The Multiprocessing Factor will be identified with a processor-oriented M/G/1
model representing the run-queue where multiprocessor interference
is associated with a staged service distribution.
\end{enumerate}
Single class workloads are assumed throughout since that will prove sufficient for the
analysis of (\ref{eqn:amdsize}) and (\ref{eqn:mpfsize}).

Just as queueing delays for elementary queues can have vastly different analytic forms
(assuming a closed analytic form exists), it would be useful to select the parametric sizing model
on the basis of the underlying queueing dynamics along the lines indicted earlier for the sizing of buffers.

\section{Multiprocessor Scalability} \label{adhoc}
We begin by briefly reviewing the conventional intuition behind the single parameter sizing models in 
(\ref{eqn:amdsize}) and (\ref{eqn:mpfsize}).

\subsection{Multiprocessor Speedup}
Amdahl's law~\cite{amdahl} is well-known and
frequently cited in the context parallel processing performance
(\cite{ware}, \cite{flatt}, \cite{karp}) where is it also known as the \emph{speedup}
(\cite{gelenbe}, \cite{patterson}).
The underlying notion is that for a fixed workload size
\footnote{\cite{gust} noted that a workload scaled to the number of processors
could recover linear behaviour under certain ideal circumstances. We shall not consider such
exceptional cases here.}
there is a fraction $\sigma \in (0, 1)$ of the workload for which the execution
time remains constant as p increases. Ultimately, this fraction dominates the
speedup function causing it to become sublinear.

In the subsequent queueing analysis, it will be more useful to use the dual
representation of processing capacity based on relative throughput or \emph{scaleup}~\cite{njgBOOK00}:
\begin{equation}
C(p) = \frac{X(p)}{X(1)} \label{eqn:relcap}
\end{equation}
(\ref{eqn:relcap}) is reflective of the motivation for selling multiprocessors that support
commercial applications. There, the goal is to accommodate incremental user growth
through the purchase of increased processor capacity while minimizing the degradation to single
user responsiveness.

Assume the number of users (N) per processor is fixed (i.e., $N / p = const.$).
Let $R_{1}$ be the mean response time experienced by N users on a single processor.
We would like to maintain the response times at $R_{1}$ but
adding another processor with N users (now 2N total users across 2 processors), we find
$R_{2} > R_{1}$ due to the multiprocessor effect.

Defining the number of completed transactions per processor as \emph{c}, the
uniprocessor throughput is $X(1) = c / R_{1}$.
For 2 processors, the throughput becomes $X(2) = 2 c / R_{2}$ where the response time $R_{2}$ with 2 processors
is sightly longer than  $R_{1}$ by a fractional amount $\sigma ~R_{1}$. In other words,
$X(2) = 2 c / (1 + \sigma) R_{1}$. For 3 processors we have $X(3) = 3 c / (1 + 2 \sigma) R_{1} $.

Generalizing to p processors, the throughput is $X(p) = p~c / R_{p}$ where
$R_{p} = R_{1} + (p - 1)~\sigma R_{1}$ accounts for the fractional increase in response time due to the
activity of users on other (p - 1) processors.
Substituting $X(1)$ and $X(p)$ into (\ref{eqn:relcap}) produces:
\begin{equation}
C(\sigma,p) =  \frac{p~c}{T_{p}} ~\frac{R_{1}}{c}  = \frac{p~R_{1}}{R_{1} ~+~ (p - 1)~\sigma~R_{1}}  \label{eqn:scaleup}
\end{equation}
which, after the elimination of $R_{1}$, is identical to (\ref{eqn:amdsize}).
The asymptotic capacity is:
\begin{equation}
\lim_{p \rightarrow \infty} C(\sigma, p) = \frac{1}{\sigma}
\end{equation}

The reason that the expressions for the speedup in (\ref{eqn:amdsize}) and the
scaleup in (\ref{eqn:scaleup}) are identical (i.e., duals of each other) follows from:
\begin{displaymath}
\frac{(1 - \sigma) / p}{\sigma} = \frac{(1 - \sigma)}{p \sigma}
\end{displaymath}
The key quantity that determines the sublinear capacity is the ratio of the
``parallel'' portion $(1 - \sigma)$ to ``serial'' portion $\sigma$ of the workload. With
respect to that ratio, it is inconsequential whether the parallel portion is
scaled down by p or the serial portion is scaled up by p. The effect on the
ratio is the same.

\subsection{Multiprocessing Factor}
The multiprocessing factor (MPF) is intended as a measure of
how much effective processor capacity is available (or lost) as more processors are added
to the backplane. 
\begin{figure}[!hbtp]
\begin{center} 
\includegraphics[bb = 0 0 288 177, scale = 0.75]{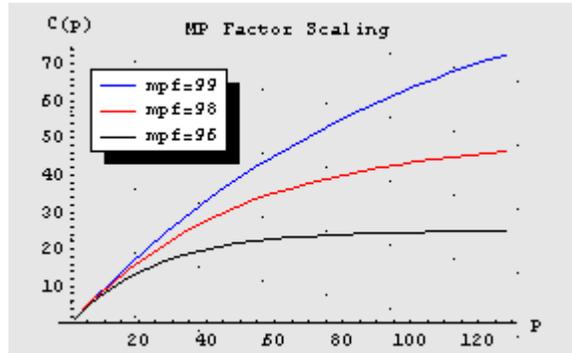} 
\caption{Different MPF factors.} ~\label{fig:coxmpf}
\end{center}
\end{figure}
Consider a workload running on a uniprocessor that has a measured throughput of
X(1) = 100 transactions per second (TPS).  When run on a dual processor the
aggregate throughput is measured as X(2) = 180 TPS. Since X(2) is less than
double X(1), this loss can be expressed as: $180 = (1 + \phi) 100$ TPS, where
the quantity $\phi = 0.8$ is the MPF. The second processor only contributes 80
percent of the capacity
\footnote{Notice that this value differs from that which would be obtained by taking the
simple arithmetic average $\frac{180}{2(100)} = 0.90$. 
}
of the first processor. Continuing along
these lines, a third processor would only be expected to contribute 80\% of the
second processor i.e., 64 TPS. The aggregate throughput being: $X(3) = X(1) +
\phi X(1)+ \phi (\phi X(1))$ = 244 TPS.

Generalizing this cumulative procedure and applying the definition in (\ref{eqn:relcap}) produces:
\begin{equation}
C(\phi, p) = 1 ~+~ \phi ~+~ \phi^{2} ~+~ \ldots ~+~ \phi^{p-1} \label{eqn:geoseries}
\end{equation}
which is equivalent to (\ref{eqn:mpfsize}) for $\phi < 1$ since it is a finite geometric sum.
The asymptotic capacity is:
\begin{equation}
\lim_{p \rightarrow \infty} C(\phi, p) = \frac{1}{1 -\phi}
\end{equation}
If $\phi = 1$ (no MPF), then
\begin{equation}
C(1, p) = \sum_{k=0}^{p}~\phi^{k} \equiv p \label{eqn:geolinear}
\end{equation}
which is a linear rising function representing ideal multiprocessor scalability.

For the purposes of comparison, (\ref{eqn:amdsize}) can also be written as a finite series
\begin{equation}
C(\sigma, p) = 1 ~+~ A_{1} ~+~ A_{2} ~+~ \ldots ~+~ A_{p - 1} \label{eqn:amdseries}
\end{equation}
where 
\begin{displaymath}
A_{i} = \frac{1 - \sigma}{1 + \sigma~(p - 1)}, ~~i = 1, 2, \ldots, (p - 1).
\end{displaymath}
Unfortunately, (\ref{eqn:amdseries}) is not a power series 
\footnote{The denominator in (\ref{eqn:amdsize}) can be expanded as a power series 
but it is an \emph{infinite} series.
}
like (\ref{eqn:geoseries}),
so the choice of scaling equation is not obvious even in a series
representation. Other ambiguities persist. (\ref{eqn:amdsize}) and (\ref{eqn:mpfsize}) could be
matched either at leading order by setting $C(\sigma, 1) = C(\phi, 1)$ as shown Figure~\ref{fig:leading}
or they could be matched asymptotically by setting $\sigma = (1 - \phi)$ as shown Figure~\ref{fig:asymptotia}.

\section{Queueing Dynamics}
In this section, we develop queueing models to resolve the ambiguities described above.

\subsection{Bus-oriented Model} \label{sec:busmodel}
The bus-oriented model comprises a closed queueing network, or Repairman
model~\cite{allen}, containing a finite number (p) of requests and K queueing
centers (K = 1 repair station and mean service demand D will be sufficient for our discussion).

\begin{figure}[!hbtp]
\begin{center}
\begin{picture}(200,100)(0,-50)
\put(-39,2){\usebox{\think}}
\put(70,-3){\usebox{\MMuno}}
\put(5,1.5){\line(1,0){20}}
\put(181,1.5){\line(0,-1){41.5}}
\put(5,1.5){\line(0,-1){41.5}}
\put(90,-33){X(p)}
\put(181,-40){\vector(-1,0){100}}
\put(105,-40){\line(-1,0){100}}
\end{picture}
\caption{Repairman model.} \label{fig:repdiagram}
\end{center}
\end{figure}
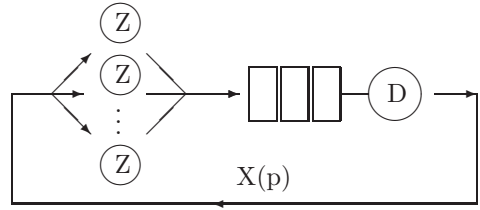

The requests can be thought of as memory references~\cite{balbo} issued by p
processors each of which executes in ``parallel'' for a mean time (Z). The
queueing center represents a ``serial'' bus or other interconnect
network~\cite{logp} by which the processors can communicate or transfer data.
Typically, we expect $D ~\ll~ Z$ to hold because the mean execution periods
should exceed the mean transit times across the bus.

System throughput (X) and communication latency (R) are related by:
\begin{equation}
X(p) = \frac{p}{R(p) + Z} \label{eqn:thru}
\end{equation}
The saturation bound $X_{max} = 1 / D$ represents the maximum throughput the
multiprocessor can achieve.

\subsubsection{Synchronous Requests} 
The worst case bound~\cite{qsp} on multiprocessor throughput ($X_{min}$) occurs
when all p processors issue synchronous communication requests. Then $R(p) =
p~D$ (maximal queueing) and (\ref{eqn:thru}) becomes:
\begin{equation}
X_{sync}(p) = \frac{p}{p D + Z} \label{eqn:amdsynch}
\end{equation}
Using the definition in (\ref{eqn:relcap}), we can use (\ref{eqn:amdsynch}) to write:
\begin{equation}
C_{sync}(p)  =  \frac{p ~(D + Z)}{p D + Z}  =  \frac{p}{p \left ( \frac{D}{D + Z} \right ) + \left ( \frac{Z}{D + Z} \right )}
\end{equation}
Rearranging terms and simplifying produces:
\begin{equation}
C_{sync}(p) = \frac{p}{(p~-~1)\left (\frac{D}{D + Z} \right ) + 1 }   \label{eqn:bound}
\end{equation}
We immediately recognize (\ref{eqn:bound}) as a version of (\ref{eqn:amdsize})
\begin{figure}[!hbtp]
\begin{center} 
\includegraphics[bb = 0 0 288 177, scale = 0.75]{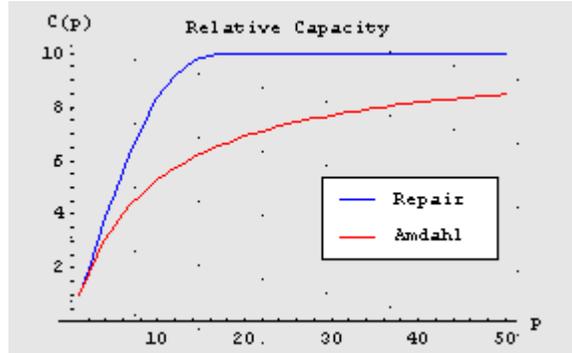} 
\caption{Capacity of the Repairman and Amdahl models.} ~\label{fig:amdrepairC}
\end{center}
\end{figure}
where the parameter $\sigma$ is now identified with the queueing parameters D
and Z via the ratio:
\begin{equation}
\sigma = \frac{D}{D + Z} \label{eqn:ratio}
\end{equation}
The range of values for $\sigma$ in (\ref{eqn:ratio}) evidently corresponds to: 
\begin{enumerate}
\item ~~$\sigma \rightarrow 0$ as $D \rightarrow 0$ (zero latency)
\item ~~$\sigma \rightarrow 1$ as $Z \rightarrow 0$ (zero execution)
\end{enumerate}
(\ref{eqn:ratio}) establishes that $C_{sync}(p)$ in (\ref{eqn:bound}) is identical to
$C(\sigma, p)$ in (\ref{eqn:amdsize}).  
Although the queue-theoretic bound (\ref{eqn:amdsynch}) on throughput is known~\cite{qsp},
its relationship to the Amdahl scaleup (\ref{eqn:scaleup}) seems not to have been discussed in the
literature.

The bus-oriented queueing model also supports an earlier conclusion \cite{njgCMG96}, \cite{njgBOOK00} that
Amdahl's law can be interpreted as representing a kind of ``broadcast protocol''
where all execution simultaneously halts while processors exchange messages
across the communication fabric.

\subsubsection{Response Times} 
The general response time for the Repairman model in Fig. (\ref{fig:repdiagram}) is given by:
\begin{equation}
R(p) = \frac{p}{X(p)} - Z  \label{eqn:rtrepair}
\end{equation}
The response characteristics for the bus-oriented model can be determined by 
substituting (\ref{eqn:amdsynch}) into (\ref{eqn:rtrepair}) and simplifying:
\begin{equation}
R_{sync}(p) = p~D \label{eqn:rtsynch}
\end{equation}
\begin{figure}[!hbtp]
\begin{center} 
\includegraphics[bb = 0 0 288 177, scale = 0.75]{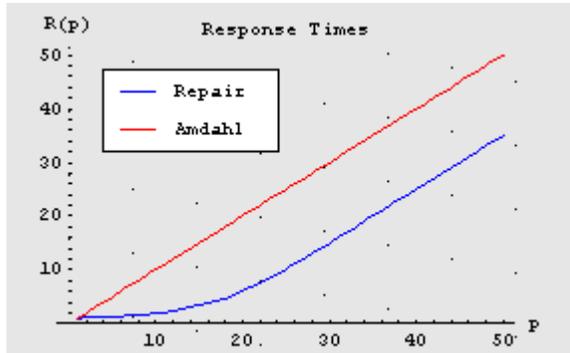} 
\caption{Response characteritics of the Repairman and Amdahl models.} \label{fig:amdrepairR}
\end{center}
\end{figure}
We see that the relative response time 
\begin{equation}
\frac{R_{sync}(p)}{R_{sync}(1)} = p 
\end{equation}
corresponding to the Amdahl bound is a linear function of p (Fig.
~\ref{fig:amdrepairR}) and independent of $\sigma$ because the system is already
in severe saturation due to synchronized queueing.

\subsection{Processor-oriented Model} \label{sec:procmodel}
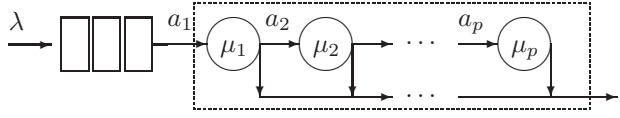
\begin{figure}[!hbtp]
\begin{center}
\begin{picture}(100,100)(0,-50)
\put(-70,10){$\lambda$}
\put(-75,-.65){\usebox{\waitline}}
\put(0,0){\usebox{\coxian}}
\end{picture}
\caption{Coxian server.} \label{fig:coxdiagram}
\end{center}
\end{figure}

The Coxian distribution~\cite{allen} represents a type of composite server  (see Fig. \ref{fig:coxdiagram})
~\cite{balbo}, \cite{qsp} with staged exponentially distributed 
service rates $\mu_{i}$ for i = 1,2, \ldots, p stages,
and probability 
\begin{displaymath}
A_{i} = \prod_{i = 0}^{p - 1}~a_{i},
\end{displaymath}
of advancing
to the $i^{th}$ server and branching probability $b_{i} $ of exiting after
the $i^{th}$ server. The next request cannot enter the service facility until
the current request has either completed all stages or exited after the $i^{th}$
stage. Consequently, there is no queueing at any of the Coxian stages.

The expected service time (first moment) is:
\begin{equation}
E\{S\} = \sum_{i = 1}^{p} ~A_{i}b_{i} \left ( \prod_{j = 1}^{i}~\frac{1}{\mu_{j}}\right ) \label{eqn:coxavg}
\end{equation}
with variance:
\begin{equation}
Var\{S\} = E\{S^{2}\} - E^{2}\{S\} \label{eqn:coxvar}
\end{equation}
where the second moment is given by:
\begin{equation}
E\{S^{2}\} = \sum_{i = 1}^{p} ~A_{i}b_{i} \left [ 
\left ( \prod_{j = 1}^{i}~\frac{1}{\mu_{j}^{2}}\right ) + \left ( \prod_{j = 1}^{i}~\frac{1}{\mu_{j}}\right )^{2} \right ] \label{eqn:coxtwo}
\end{equation}
A well-known~\cite{allen} special case is the Erlang-k distribution where all 
\mbox{$\mu_{i} = \mu$} and all \mbox{$b_{i} = 0$} and
squared coefficient of variation $C^{2}\{S\} = 1/ p$~\cite{harrison}.

\subsubsection{Uniform Coxian}
The special case of interest to us is, \mbox{$\mu_{i} = \mu$}, 
\mbox{$a_{i} = \phi$} with \mbox{$a_{p} = 0$} and \mbox{$b_{i} = 1 - \phi$} with \mbox{$b_{0} = 0$}. 
We shall refer to this as a \emph{uniform Coxian distribution}.
Then (\ref{eqn:coxavg}) reduces to:
\begin{equation}
E\{S\} = \frac{1}{\mu} + \frac{\phi}{\mu} + \frac{\phi^{2}}{\mu}  + \dots + \frac{\phi^{p-1}}{\mu} \label{eqn:stime}
\end{equation}

Using Little's law $U = \lambda E\{S\}$, (\ref{eqn:stime}) can be rewritten as: 
\begin{equation}
U(\phi, p) = \frac{\lambda}{\mu}~ \left ( \frac{~1 - \phi^{p}}{1 -\phi} \right ) \label{eqn:util}
\end{equation}
Since $U(\phi, p) \geq 1$, (\ref{eqn:util}) represents 
the \emph{total} utilization of the uniform Coxian service facility.
It is bounded above by 
\begin{displaymath}
U(\phi, p) \leq \frac{\rho}{1 - \phi} 
\end{displaymath}
where $\rho = \lambda / \mu$.
Moreover, (\ref{eqn:mpfsize}) can now be expressed in terms of (\ref{eqn:util}) as:
\begin{equation}
\rho~C(\phi, p) = U(\phi, p) \label{eqn:new}
\end{equation}
Hence, the MPF capacity model presented in section \ref{adhoc} 
can also be interpreted as the total utilization of a p-stage
uniform Coxian server.
For a single-stage server p = 1 and (\ref{eqn:new}) reduces to $U(\phi, 1) = \rho$, as expected.

In this queueing model, the finite geometric 
\footnote{The geometric series in (\ref{eqn:stime}) should not be confused with
the geometrically distributed probability $p_{k} = \rho^{k}(1 - \rho)$ of finding k customers 
in an M/M/1 queue.}
series in (\ref{eqn:geoseries}) arises from the
\emph{branching} process within the service center (not the arrivals
process). 
This branching represents the loss of service after some number of processor cycles due to system overhead.
The total utilization $U(\phi, p)$ corresponds to the average impact of that loss.

\subsubsection{Response Times} 
\begin{figure}[!hbtp]
\begin{center} 
\includegraphics[bb = 0 0 288 177, scale = 0.75]{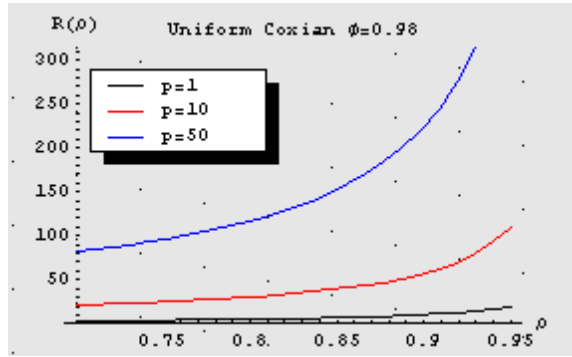} 
\caption{Response time $R(\rho)$ as a function of $\rho$.} \label{fig:coxRvRho}
\end{center}
\end{figure}

The corresponding response times (Fig.~\ref{fig:coxRvRho}) for the uniform
Coxian model can be calculated as an M/G/1
queue using the Pollacek-Khintchine formula~\cite{harrison}:
\begin{equation}
R(\phi, \rho) = E\{S\} \left [ 1 ~+ ~\frac{\rho~(1 +  C_{\phi}^{2}\{S\})}{2~(1 - \rho)} \right ] \label{eqn:pk}
\end{equation}
where the squared coefficient of variation
\begin{displaymath}
C_{\phi}^{2}\{S\} = \frac{Var\{S\}}{E^2\{S\}} 
\end{displaymath}
is defined in terms of (\ref{eqn:coxavg}) and (\ref{eqn:coxtwo}) and lies in the range
$1 / p < C_{\phi}^{2}\{S\} < 1$. As expected, the uniform Coxian model represents a
hypoexponential server. The variance in the service time is smaller than it would be
for an M/M/1 queue.

\begin{figure}[!hbtp]
\begin{center} 
\includegraphics[bb = 0 0 288 177, scale = 0.75]{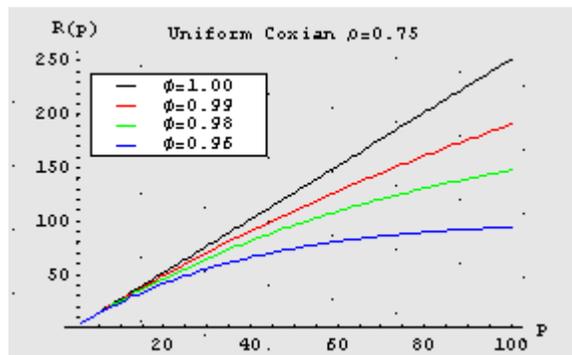} 
\caption{Response time $R(p)$ vs. the number of stages for $\rho=0.75$.} ~\label{fig:coxRvStage}
\end{center}
\end{figure}

The response time (\ref{eqn:pk}) is plotted as a function of $\rho$ in figure
(\ref{fig:coxRvRho}) for a fixed value of $\phi$. It has the typical
characteristic expected of an open class queue. With only a single stage and
probability 1 of advancement ($\phi = 0.98$) $R(\rho)$ is close to an M/M/1
queue since $C_{\phi}^{2}\{S\} \sim 1$. As more stages are added, the response
time at any load $\rho$ increases as shown by the curves for p = 10 and p = 50.
Note, however, that the progressive increase at that load becomes smaller as the
number of stages increases. This effect can be seen more clearly in figure
(\ref{fig:coxRvStage}) which shows response times plotted as a function of p for
a fixed load $\rho = 0.75$ up to 100 stages representing a large-scale
multiprocessor. A surprising feature, for modeling multiprocessors, is that the
response time characterisitcs are sublinear for all $\phi < 1$.
Contrast this with the response time characterisitcs in figure (\ref{fig:amdrepairR}).

Only for the special case $\phi = 1$ (Erlang-p), does the response time increase
linearly because there, all the processor work is accounted
for i.e., $U(\phi, p) = p\rho$. That case, however, is tantamount to linear
scalability in (\ref{eqn:geolinear}) which ignores the MP effect and is therefore
of little value for multiprocessor sizing.

The queue-theoretic attributes of the MPF model can be summarized as follows:
\begin{enumerate}
\item Only one request at a time can enter the Coxian server.
\item Multiprocessor overhead is treated as a probabilistic loss of work.
\item Processor utilization due to the MP effect is unaccounted for.
\item Service periods are hypo-exponential.
\item Response times become sublinear with an increasing number of processors.
\end{enumerate}
These characteristics appear counter-intuitive as a model of multiprocessor scalability.

\section{Conclusions}
Based on our queueing analysis of these multiprocessor models we are now
in a position to say something about the applicability of the bus-oriented
(Amdahl) model defined by (\ref{eqn:amdsize}) and the server-oriented model (MPF)
model defined by (\ref{eqn:mpfsize}).

If matched at small processor configurations, both capacity models are essentially
indistinguishable when fitted to benchmark data. As configurations become
larger, however, the MPF model becomes pessimistic relative to the Amdahl model.
This appears contradictory when we recall that Amdahl scaling corresponds
to the worst-case bound of the more constrained closed queueing model.

Capacity scaling for the bus-oriented (Amdahl) model in section
\ref{sec:busmodel} is an explicit function of the system throughput X(p).
Response times for bus-oriented (Amdahl) model will have the classic
``hockey-stick'' shape due to the negative feedback effects of a finite number
of requests in the closed queueing network. Such response time curves are
associated with the constraint that no more than one bus request per processor
can be outstanding. Utilizations of both the bus, U(p), and the processors, Z.X(p),
are accounted for explicitly.

Based on the discussion in section \ref{sec:procmodel}, the relative capacity for
server-oriented model (MPF) model is equivalent to the total utilization U(p) of
a p-stage Coxian server. For a given value of $\phi$, the total utilization
becomes sublinear with increasing stages because the likelihood diminishes that
a request will visit all stages. In this model, multiprocessor overhead is
treated as a loss of serviceable work.

Considered as an M/G/1 queue, the multiprocessor is represented as a single
Coxian server with processor interference accounted for by the variance in the
service period. That only one request can enter the Coxian server at a time is
already unrealistic for a model of a multiprocessor but a variance in the
service periods that is \emph{less} than an exponential server (i.e., \emph{hypo}exponential), 
seems contradictory to expectations for a model of the
multiprocessor effect. 
M/G/1 queues with $C^{2}\{S\} \gg 1$ (i.e., non-Coxian) have been
used to model disk storage and token ring networks~\cite{harrison}, however, we need
the Coxian stages to account for the geometric series in (\ref{eqn:mpfsize}). 

A \emph{hyper}exponential Coxian would also produce higher variance $C^{2}\{S\} > 1$ 
in the service periods but it is well known (\cite{allen}, \cite{harrison}) that just a few
parallel stages are sufficient for that and thus p could no longer be associated
explicitly with the number of processors. Moreover, and hyper-exponential Coxian
does not produce a mean service time that has the geometric series required to
account for (\ref{eqn:mpfsize}).

The response time for the Coxian server model becomes sublinear as the
processor configuration is expanded. This is unlikely to be seen in benchmark
measurements of real multiprocessors. Such an unphysical effect follows from the
fact that the utilization due to lost processor work is unaccounted for in the
Coxian model. In reality, one expects multiprocessor overhead to be accrued as
processor kernel time rather than processor user time. The total processor
utilization is the sum of both contributions but the uniform Coxian server does not
account for kernel time in the workload.

Finally, we suggest that neither of the models considered here is truly sufficient
as a general model of multiprocessor scalability. Elsewhere, we have already proposed
a two-parameter model~\cite{njgCMG93}:
\begin{equation}
C(\alpha, \beta, p) = \frac{p}{1 + \alpha~[(p - 1) + \beta~p~(p - 1)]} \label{eqn:gunther}
\end{equation}
in which the $\alpha$ parameter is
identified with queueing delays and the $\beta$ parameter with additional delays due to 
pairwise coherency \cite{lambda}
mismatches \cite{njgBOOK00}. The latter induces retrograde throughputs 
\mbox{C($\alpha$, $\beta$, p)} $\rightarrow$ 1/p as p $\rightarrow \infty$
that are indeed seen in multiprocessor capacity measurements~\cite{njgBOOK01}. 
Retrograde throughput cannot be modeled parametrically using either 
(\ref{eqn:amdsize}) or (\ref{eqn:mpfsize}) nor can it be represented using conventional
queueing theory without the introduction of load-dependent servers such that $E\{S\} \sim 1/p$.
In the limit where coherency penalties vanish ($\beta = 0$), (\ref{eqn:gunther})
reduces to the Amdahl model (with $\alpha = \sigma$) in (\ref{eqn:amdsize}). 
As we have demonstrated here, Amdahl's law has a natural physical
interpretation as synchronous queueing within a Repairman model.
The two-parameter function (\ref{eqn:gunther}) can be viewed as a load-dependent extension
of that queueing dynamics.

Although we have been able to show that the MPF scaling equation (\ref{eqn:mpfsize}) belongs
to an M/G/1 queueing model with a load-dependent Coxian server, that
load dependence is not of the correct type for modeling multiprocessor overhead
because it gives rise to unphysical effects. We therefore caution against its use for large-scale 
multiprocessor servers.

\pagebreak

\bibliography{BksCS,BksPerf,BksStat,NJG,PapPerf,PapURL,GeoCox}

\begin{thebibliography}{10}

\bibitem{amdahl}
G.~Amdahl.
\newblock ``{V}alidity of the single processor approach to achieving large
  scale computing capabilities''.
\newblock {\em {P}roc. {AFIPS} Conf.}, 30:483--485, Apr. 18--20 1967.

\bibitem{patterson}
J.~L. Hennessy and D.~A.Patterson.
\newblock {\em Computer Architecture: A Quantitative Approach}.
\newblock Morgan Kaufmann, 2nd. edition, 1996.

\bibitem{gelenbe}
E.~Gelenbe.
\newblock {\em Multiprocessor Performance}.
\newblock Wiley, 1989.

\bibitem{ware}
W.~Ware.
\newblock ``{T}he ultimate computer''.
\newblock {\em IEEE Spectrum}, pages 89--91, March 1982.

\bibitem{artis}
H.~P. Artis.
\newblock ``{Q}uantifying multiprocessor overheads''.
\newblock {\em Proc. CMG Conference}, pages 363--365, 1991.

\bibitem{cockwalk}
A.~Cockcroft and W.~Walker.
\newblock {\em Sun Blueprints: Capacity Planning for Internet Services}.
\newblock Prentice--Hall, New Jersey, 2000.

\bibitem{mcgall}
J.~W. McGalliard.
\newblock ``{C}ase study of table-top sizing with workload-specfic estimates of
  the multiprocessor effect''.
\newblock {\em Proc. CMG Conference}, Nashville, Tennesee:208--217, December
  1995.

\bibitem{gsa}
J.~L. Fitch.
\newblock ``{A} guide for performance and capability validation''.
\newblock Document KMP--94--2--P, U.S. General Services Administration, 1993.

\bibitem{ibm}
W.~Bitner.
\newblock ``{VM/ESA} greater {N}--way thoughts''.
\newblock \tt \footnotesize
  http://www.vm.ibm.com/devpages/bitner/presentations/vmnway.html
  \#Section\_0.6\rm\normalsize, October 1996.

\bibitem{hitachi}
Hitachi Data~Systems Inc.
\newblock ``{H}itachi unleashes new series of e--{B}usiness mega servers with
  world's highest performance and availability''.
\newblock \tt \footnotesize http://al.hds.com/news/000207.html\rm\normalsize,
  2000.

\bibitem{unisys}
David Floyer.
\newblock ``{T}he performance and price/performance of current mainframe
  systems''.
\newblock \tt \footnotesize
  http://www.unisys.com/hw/servers/clearpath/16222.htm\rm\normalsize, May 1998.

\bibitem{lsprRB}
IBM.
\newblock ``{L}arge systems performance reference''.
\newblock \tt \footnotesize
  http://www-1.ibm.com/servers/eserver/zseries/lspr/\rm\normalsize, February
  2002.

\bibitem{spec}
System Performance Evaluation~Corporation SPEC.
\newblock ``{CPU2000} benchmark''.
\newblock \tt \footnotesize http://www.spec.org/osg/cpu2000/\rm\normalsize,
  2000.

\bibitem{tpcc}
Transaction Processing Performance~Council TPC.
\newblock ``{TPC--C} benchmark''.
\newblock \tt \footnotesize http://www.tpc.org/tpcc/\rm\normalsize, 2000.

\bibitem{tpcw}
Transaction Processing Performance~Council TPC.
\newblock ``{TPC--W} benchmark''.
\newblock \tt \footnotesize http://www.tpc.org/tpcw/\rm\normalsize, 2000.

\bibitem{njgCMG96}
N.~J. Gunther.
\newblock ``{U}nderstanding the {MP} effect: Multiprocessing in pictures''.
\newblock {\em Proc. CMG Conference}, San Diego, California:957--968, December
  1996.

\bibitem{njgBOOK00}
N.~J. Gunther.
\newblock {\em The Practical Performance Analyst}.
\newblock iUniverse.com, Inc., Lincoln, Nebraska, {P}rint-{O}n-{D}emand
  edition, 2000.

\bibitem{flatt}
H.~P. Flatt.
\newblock ``{F}urther results using the overhead model for parallel systems''.
\newblock {\em IBM J. Res Develop.}, 35(5/6):721--726, 1991.

\bibitem{karp}
A.~H. Karp and H.~P. Flatt.
\newblock ``{M}easuring parallel processor performance''.
\newblock {\em Comm. ACM}, 33(5):539--543, 1990.

\bibitem{gust}
J.~L. Gustafson.
\newblock ``{R}eevaluating {A}mdahl{'}s law''.
\newblock {\em Comm. {ACM}}, 31(5):532--533, 1988.

\bibitem{allen}
A.~O. Allen.
\newblock {\em Probability, Statistics, and Queueing Theory with Computer
  Science Applications}.
\newblock Academic Press, San Diego, 2nd. edition, 1990.

\bibitem{balbo}
M.~Ajmone-Marsan, G.~Balbo, and G.~Conte.
\newblock {\em Performance Models of Multiprocessor Systems}.
\newblock MIT Press, Cambridge, Mass., 1990.

\bibitem{logp}
D.~E. Culler, R.~M. Karp, D.~Patterson, A.~Sahay, E.~E. Santos, K.~E. Schauser,
  R.~Subramonian, and T.~Eicken.
\newblock ``{L}og{P}: A practical model of parallel computation''.
\newblock {\em Comm. ACM}, 39(11):79--85, November 1996.

\bibitem{qsp}
E.~D. Lasowska, J.~Zahorjan, G.~S. Graham, and K.~C. Sevcik.
\newblock {\em Quantitative System Performance: Computer System Analysis Using
  Queueing Network Models}.
\newblock Prentice--Hall, Engelwood Cliffs, 1984.

\bibitem{harrison}
P.~G. Harrison and N.~M.Patel.
\newblock {\em Performance Modelling of Communication Networks and Computer
  Architectures}.
\newblock Addison--Wesley, Wokingham, U. K., 1993.

\bibitem{njgCMG93}
N.~J. Gunther.
\newblock ``{A} simple capacity model for massively parallel transaction
  systems''.
\newblock {\em Proc. CMG Conference}, San Diego, California:1035--1044,
  December 1993.

\bibitem{lambda}
S.~Cho and G.~Lee.
\newblock ``{R}educing cache coherence overhead in shared--bus
  multiprocessors''.
\newblock {\em Proc. EURO--PAR'96}, August 1996.

\bibitem{njgBOOK01}
N.~J. Gunther.
\newblock ``{P}erformance and scalability models for a hypergrowth e-{C}ommerce
  {W}eb site''.
\newblock In R.~Dumke, C.~Rautenstrauch, A.~Schmietendorf, and A.~Scholz,
  editors, {\em Performance Engineering: State of the Art and Current Trends},
  volume \# 2047, pages 267--282. Springer--Verlag, Heidelberg, 2001.

\end{thebibliography}
\bibliographystyle{unsrt}

\end{document}